# Engineering the magnetic anisotropy of atomic-scale nanostructure under electric field

Wanjiao Zhu[1], Hang-Chen Ding[1], Wen-Yi Tong[1], Shi-Jing Gong[1], Xiangang Wan[2], Chun-Gang Duan[1, 3]

[1] Key Laboratory of Polar Materials and Devices, Ministry of Education, East China Normal University, Shanghai 200241, China

[2] Department of Physics and National Laboratory of Solid State Microstructures, Nanjing University, Nanjing 210093, China

[3] National Laboratory for Infrared Physics, Chinese Academy of Sciences, Shanghai 200083, China

**Abstract**

Atomic-scale magnetic nanostructures are promising candidates for future information processing devices. Utilizing external electric field to manipulate their magnetic properties is an especially thrilling project. Here, by careful identifying different contributions of each atomic orbital to the magnetic anisotropy energy (MAE) of the ferromagnetic metal films, we argue that it is possible to engineer both the MAE and the magnetic response to the electric field of atomic-scale magnetic nanostructures. Taking the iron monolayer as a matrix, we propose several interesting iron nanostructures with dramatically different magnetic properties. Such nanostructures could exhibit strong magnetoelectric effect. Our work may open a new avenue to the artificial design of electrically controlled magnetic devices.

# I. Introduction

Electric field control of magnetism, due to its overwhelming advantage in energy consumption and compatibility with modern semiconductor technology, recently has attracted intensive researches [1-10]. The key problem in this field is how exactly the applied field could influence the magnetic properties, especially the magnetic anisotropy energy (MAE), of the system, or what kind of structure or system will have large magnetic response to the external electric field. For ferromagnetic metal films, which are widely used in modern magnetic recording technology, it is generally believed that the electric field control of MAE is made possible due to the fact that the accumulated spin charge can modify the occupation states in $3d$ orbitals at the surface and consequently change the MAE [7, 11, 12]. Experimentally, impressive progresses have been made to use applied field (voltage) to affect the MAE of ferromagnetic metal films [4, 8-10, 13, 14].

However, there still exist serious problems in the electric direct control of MAE in the ferromagnetic metal systems. Previous reports have shown that the MAE magnetoelectric coefficient is relatively limited [7, 15], which is in the magnitude of $\mu J/V\ m^2$. Thus it is necessary to apply an extreme high electric field to change the MAE of the system in a broad range and accomplish the magnetization reversal, which may restrict the research and development of related devices. In addition, it is still not very clear that which orbital has made the main contribution in the electric manipulation of MAE, which is important to achieve higher MAE response to the external field. Different magnetic systems have dramatically different responses to the external electric field, and the magnetoelectric coefficients could have different sign for, at least appears to be, similar systems [7, 9, 10, 15].

On the other side, magnetic nanostructures recently become the research focus due to their fascinating material properties emerging from the low-dimensional character of the system [16, 17]. Furthermore, both theoretical [18, 19] and experimental (spin-polarized scanning tunneling microscopy) [20] studies demonstrate that external electric field could be used to manipulate the magnetic properties of the atomic-scale nanomagnets. As magnetic nanostructures provide rich opportunity to MAE tailoring, it is therefore the aim of our present study to find the clue to tune the electric field response of magnetic systems.

This paper is organized as follows. In the next section we introduce the method used in the analysis of MAE of ferromagnetic films. Particularly we emphasis a new approach to identify the physical contribution of specific atomic orbital. Then in Section III we first taking Fe monolayer as a protype system to demonstrate the orbital dependent analysis of the magnetic anisotropy of the system. Then we propose several Fe/Cu magnetic nanostructures and apply electric field to study their magnetic response. We find that different magnetic nanostructures demonstrate significantly different magnetoelectric effect, which suggests they have great flexibility in electric field control of magnetism.

**II. Method**

In order to understand the mechanism of the direct electric modulation of MAE in the ferromagnetic metals, we should consider the spin-orbital coupling (SOC), which is the main origin of the magnetocrystalline anisotropy [21]. For discussing the relationship between the electron occupation and MAE including the SOC, one direct method is to consider a perturbative formula as follows [22, 23]:

$$E_x - E_z \sim \xi^2 \sum_{o,u} \frac{|\langle o|l_z|u\rangle|^2 - |\langle o|l_x|u\rangle|^2}{\varepsilon_u - \varepsilon_o} \quad (1)$$

where $u$ and $o$ represent the unoccupied and occupied states, respectively, $l_{z(x)}$ is the $z(x)$ component of the angular momentum operators. $\varepsilon_u$ and $\varepsilon_o$ are the energy of unoccupied and occupied states. $\xi$ is the spin-orbital coupling parameter. We should point out that very recently Barnes *et al.* proposed that Rashba SOC may also contribute to the MAE modification due to the applied electric field [24]. Such mechanism, however, may not play important roles in 3$d$ ferromagnetic metal films which only show weak Rashba effect.

Eq. (1), though appearing in a simple form, in fact is difficult to be used in analyzing the contribution of each atomic orbitals. Only in some special cases, it is relatively easier to identify the origin of MAE of the system. Examples are typical ferromagnetic metals, such as Fe, Co, Ni, in which the majority spin states in the 3$d$ orbitals are fully occupied and almost have no influence on the orbital angular momentum and MAE. Then we can use van der Laan's formula [23] to trace the reason causing the MAE change of the system [15, 25]. For other magnetic systems, the MAE analysis is generally not straightforward [26].

Recently, we have developed a strategy called orbital selective external potential (OSEP) method [27, 28]. The spirit of this method is to introduce a special external potential on specific atom of a system. Distinguishing from the realistic potential, this potential is orbital dependent, i.e. can only be felt by specifically assigned atomic orbital. By introducing one projector operator $|inlm\sigma\rangle\langle inlm\sigma|$, the new Hamiltonian can be written as:

$$H^{\text{OSEP}} = H^0_{\text{KS}} + |inlm\sigma\rangle\langle inlm\sigma|V_{\text{ext}} \qquad (2)$$

where $H^0_{\text{KS}}$ is the unperturbed Kohn–Sham Hamiltonian, and $V_{\text{ext}}$ is the applied external potential energy. Index $i$ represent the atom site, while $n$, $l$, $m$ and $\sigma$ are the orbital index. Through the OSEP method, we can effectively shift the energy level of the specified orbitals and consequently manipulate the electron occupation number on the orbitals, which provides a direct way to study the influence of electron occupation on the MAE of the ferromagnetic system.

To demonstrate the power of the OSEP method, we first calculate the MAE of one-layer Fe atom system ($a$=2.87 Å, the optimized lattice constant of Fe). Here the OSEP method is implemented in the Vienna *ab initio* simulation package (VASP) [29] to investigate the relationship between the electron occupation number in 3$d$ orbitals and MAE. The exchange–correlation potential is treated in the generalized gradient approximation. We use the energy cutoff of 500 eV for the plane wave expansion of the PAWs and a 10×10×1 Monkhorst–Pack grid for $k$-point mesh both in the structural relaxation and the self-consistent calculations. All the structural relaxations are performed in the absence of electric field until the force on every atom becomes less than 1meV/Å. The MAE is defined as $\Delta E = E_{/\!/} - E_\perp$ [30], where $E_{/\!/}$ and $E_\perp$ represent the total energies with the magnetization along the in-plane and perpendicular to the plane direction, respectively, and the easy axis with a positive MAE is along the $c$ axis. The convergences of MAE calculation over both cutoff energy and $k$-point sampling have been tested, as did in previous researches [7, 25]. The external electric field is introduced by planar dipole layer method [31].

## III. Results and discussion

As pointed out above, the majority-spin $d$ bands of monolayer Fe film are nearly fully occupied. We therefore only consider the effect of shifting the minority-spin $d$ bands. Table I shows the so obtained results, where $V_i$ ($i=xy, yz, z^2, xz, x^2-y^2$)= -0.5 eV is the energy shift of $d$ orbital, and the negative value of $V$ means shifting the orbital energy level downward and generally will increase the occupation of the corresponding electron state. We note that as the minority $d$ spin states are unfilled, the shift of one of them will inevitably affect the occupations of other $d$ states as well as the Fermi level. The change of the minority spin electron occupation number ($\Delta n$) due to the energy shifting is also listed in Table I. It is obvious that the shift of $d_{z^2}$ orbital induce the largest change in MAE, which is about -1.49 meV/atom and results in a strong in-plane magnetization.

In order to show how exactly the MAE changes with the increasing occupation in $d_{z^2}$ orbital, we show in Fig. 1 the curve of the relationship between the applied energy shift $V_{z^2}$ and the MAE. It can be seen that MAE decreases monotonically with the energy downshift and therefore the occupation increase of $d_{z^2}$ state. Especially around the point of $V_{z^2} = -0.3$ eV, the MAE experiences the most dramatic change and even becomes negative when additional energy shift is applied. From the inset plot of the partial density of states (PDOS) of $d_{z^2}$ in Fig. 1, we find that the applied external potential indeed pushes the energy level of $d_{z^2}$ downwards and results in significant change in the occupation of $d_{z^2}$ orbital. It also effectively raise the other 3$d$ minority-spin states near the Fermi level (comparing the red solid curve and shaded plots). Therefore, we now assure that the occupation variation in the minority spin channel of the 3$d_{z^2}$ orbital can bring about a critical effect on the MAE of the Fe monolayers and similar systems.

For situation as simple as Fe monolayer film, the above phenomenon is actually easy to be provided with an analytic explanation. From the nonvanishing matrix of the $l_z$ and $l_x$ operators [22], we find that the $d_{z^2}$ orbital has some peculiarities distinguishing from other 3$d$ orbitals. In the $l_z$ operator matrix, all the matrix elements involving the $d_{z^2}$ orbital are zero. Meanwhile, in the $l_x$ operator matrix, there exists one nonvanishing element $\langle z^2 | l_x | yz \rangle$ related to the $d_{z^2}$ orbital. Thus it indicates that the change of the electron occupation in $d_{z^2}$ orbital has little influence on the term $|\langle o|l_z|u\rangle|^2$, but it will induce a pronounced modification in the term $|\langle o|l_x|u\rangle|^2$ in Eq. (1). As for other $d$ orbitals, i.e. the $d_{xy}$, $d_{xz(yz)}$ and $d_{x^2-y^2}$, the involved nonzero matrix elements appear both in the $l_z$ and $l_x$ operators. So their change in the electron occupations will affect both the terms $|\langle o|l_z|u\rangle|^2$ and $|\langle o|l_x|u\rangle|^2$, and will generally only lead to a modest change to the MAE according to Eq. (1). From Table I, we can see that the increase of $n_{z^2}$ comes with the decrease of $n_{xz(yz)}$. In the unperturbed system, the $d_{z^2}$ state is almost unoccupied, whereas for $V_{z^2}$ =-0.5 eV $d_{xz(yz)}$ states become almost unoccupied. From the occupation change shown in Table I, we can easy to see that the occupation increase of $d_{z^2}$ state is essentially due to the electron transfer from $d_{xz}$ and $d_{yz}$ states, and therefore $n_{z^2}$ is about twice of the original $n_{yz}$. This will result in about four times enhancement of the contribution of nonzero element $\langle z^2|l_x|yz\rangle$ to the term $|\langle o|l_x|u\rangle|^2$ in Eq. (1). This change is so big that the average value of $l_x$ (~0.13) becomes larger than $l_z$ (~0.10), and the MAE changes its sign. We therefore conclude that the $d_{z^2}$ electron occupation plays the most important role in the MAE of monolayer of Fe, and could be used to tune the MAE.

In the above analysis, with the help of the OSEP method, we shift the energy states of each $d$ orbitals to show their influences on the MAE of the monolayer Fe film. But how can we physically modify the electron occupation of $d$ orbital?

It is known that the neighboring atoms, which essentially form the local molecular field, could affect the energy states and consequently the electron occupations of certain ions. We therefore consider Fe nanostructures, like Fe dimer，trimer and tetramer, or other different geometrically different systems to study the influence of the local structure to the MAE of the magnetic film. Fig. 2 shows six typical different structures in our study. Here structure (a) is normal square alignment of the Fe atoms. Structure (b) is the separated square alignment, like Fe tetramer. Structure (e) is designed as the hexagon graphene-like shape, and structure (d) is Fe dimer structure with a parallel and perpendicular arrangement. All these four structures have relatively high symmetry along the in-plane direction. Whereas structure (c) and (d) are Fe trimers with different in-plane alignment, whose in-plane symmetry is further reduced.

Then we calculate the MAE of these different structures, and the results are listed in Table II. Here $n_{z^2}$ represents the averaged occupation number of the minority spin channel in $d_{z^2}$ orbital of Fe atoms. From Table II, we find that comparing with structure (a), structures with a low in-plane symmetry, i.e. the triangle structure (c) and (d), correspond to a significant increase in $n_{z^2}$ and lead to a negative value in MAE, indicating the magnetic easy axis turns from perpendicular to in-plane orientation. Even in systems with higher in-plane symmetry, the positive MAE decreases under an increased $n_{z^2}$, due to the same reason explained in above analysis.

Based on above understanding, we then move our topic to the electric field control of the MAE of the ferromagnetic nanostructures. As we now realize that the MAE of monolayer Fe is highly sensitive to $n_{z^2}$, then it is natural to think that a structure with its $n_{z^2}$ easy to be modified by the external electric field or related effect will have a large MAE change under the influence of the field. As we know the surface magnetoelectric effect in ferromagnetic metal film is mainly due to the spin-dependent screening [7], therefore we expect a large slope of $3d_{z^2}$ state at the Fermi level will result in significant field induced MAE change.

We then explore the effect of the electric field on the MAE of the designed monolayer Fe systems in Fig. 2. This is done by considering these Fe nanostructures grown on the substrate of 4 layers Cu, i.e. the Fe/Cu heterostructure where the substrate Cu layer keeps its original fcc alignment structure, and calculate the MAE under different external electric fields [7, 15]. Here we define the MAE related magnetoelectric coefficient $\beta_s$ as $\beta_s=\Delta K_u/E$, where $\Delta K_u$ is the change of MAE value per area [15].

The obtained results indeed confirm our prediction. In Fig. 3(a), we show the results of three Fe/Cu systems, i.e with the Fe layer of normal, triangle and hexagon graphene-like alignment structures, which have been shown in Figs. 2(a), 2(d) and 2(e), respectively. A linear relationship between the electric field and the MAE is well kept for all three systems. Note that here the electric field manipulated the MAE which is mainly from the contribution of magnetocrystalline anisotropy, and the shape anisotropy from the change of alignment structure is neglected. We also find that comparing with result of the Fe/Cu system of normal square Fe alignment [the fitting line for the black square scatters in the Fig. 3(a)], the slope of the triangle Fe alignment (red circle) is clearly smaller, indicating that it has a weaker MAE related magnetoelectric effect. Meanwhile, the electric tuning of MAE in Fe/Cu with the graphene-like Fe alignment (blue triangle) is very close to that of the normal square Fe alignment. Taking the surface area per Fe atom occupied into account, the MAE related magnetoelectric coefficient $\beta_s$ are about 1.1, 0.18 and $1.0 \times 10^{-9}$ erg/V cm for normal, triangle and graphene-like alignment structures, respectively.

The behaviors of the MAE under electric field of different Fe/Cu systems exactly agree with our above analysis about the 3 $d_{z^2}$ occupation of the in the minority spin channel of the surface Fe atom. Fig. 3(b) displays the PDOS of the Fe 3 $d_{z^2}$ orbital in the three Fe/Cu systems. We find that due to the charge transfer between Fe and adjacent Cu atoms, the Fermi level now crosses the Fe 3 $d_{z^2}$ state of these Fe/Cu structures, which is significantly different from the case of pure Fe monolayer. However, the one with normal Fe alignment (shaded plot) and that with hexagon alignment (blue straight line) have much larger densities and slopes at the Fermi level than those of the triangle Fe alignment Fe/Cu system. Consequently, the former two have similar magnetoelectric coefficient $\beta_s$, yet $\beta_s$ of the latter is much smaller, considering that it also has much larger surface area per Fe atom.

Finally, to further verify our theory, we again utilize the OSEP method to shift the $3d_{z^2}$ minority spin states of the surface Fe atom in the normal Fe alignment Fe/Cu system to tune the 3 $d_{z^2}$ positions to have larger MAE response to the external field. The result for $V_{z^2}$ = -0.3 eV is shown in the inset of Fig. 3(a). Note that the MAE value of OSEP becomes negative mainly due to the largely increased $3d_{z^2}$ minority spin occupation [see the dash dot line in Fig. 3(b)]. We then see it clearly that the slope of the MAE versus $E$ curve increased dramatically. The corresponding MAE coefficient $\beta_s \approx 2.3 \times 10^{-9}$ erg/V cm is more than twice of the original one. This again demonstrate that the MAE response to the external field could be enhanced and for the above studied systems it is $3d_{z^2}$ state who plays the most important role.

## IV. Conclusions

Based on above systematic researches, we conclude that it is possible to tune the MAE, as well as the MAE change under the external field, of atomic-scale magnetic nanostructures. Physically this can be achieved by structure engineering [32, 33], surface decoration, heterostructure, doping, gate voltage control [34] or surface charge injection [35, 36], *etc*. The spirit of such strategy is to find the specific atomic orbital which contributes most to the MAE, then tune its occupation and the slope of the occupation versus the energy. In some critical cases where the occupation has abrupt changes, e.g. phase transition, one may get huge MAE change due to applied electric field. We believe magnetic nanostructures, due to their various forms, are the best candidates for such studies. We hope that this work is helpful to understand the mechanism of electric field control of MAE in ferromagnetic metal systems and achieve an applicable effect on MAE manipulation in future research.

**Acknowledgments**

This work was supported by the 973 Programs No. 2014CB921104, 2013CB922301, the NSF of China (No. 61125403), NSF of Shanghai (No. 14ZR1412700), Program of Shanghai Subject Chief Scientist. Computations were performed at the ECNU computing center.

**Tables**

TABLE I. Calculated MAE (in unit of meV/atom) and change of minority spin electron occupation number $\Delta n$ under the external potential effect in monolayer Fe atom. Here $V_i$ ($i=xy, yz, z^2, xz, x^2-y^2$) represents the energy shift of Fe $3d$ orbitals and is given in unit of eV. The original minority spin electron occupation number $n_i$ ($i=xy, yz, z^2, xz, x^2-y^2$) is also listed as a reference.

| $V_{OSEP}$ (eV) | MAE (meV/atom) | | $n_{xy}$ | $n_{xz(yz)}$ | $n_{z^2}$ | $n_{x^2-y^2}$ |
|---|---|---|---|---|---|---|
| $V=0$ | 0.89 | | 0.201 | 0.366 | 0.029 | 0.495 |
| $V=-0.5$ | | $\Delta$ MAE (meV/atom) | $\Delta n_{xy}$ | $\Delta n_{xz(yz)}$ | $\Delta n_{z^2}$ | $\Delta n_{x^2-y^2}$ |
| $V_{xy}$ | 1.23 | 0.34 | 0.090 | -0.012 | 0.001 | -0.008 |
| $V_{xz(yz)}$ | 0.33 | -0.56 | 0.002 | 0.338 | 0.050 | -0.030 |
| $V_{z^2}$ | -0.60 | -1.49 | 0.012 | -0.367 | 0.803 | -0.029 |
| $V_{x^2-y^2}$ | 1.23 | 0.34 | -0.007 | -0.050 | -0.001 | 0.157 |

TABLE II. Calculated MAE and average electron occupation number $n_{z^2}$ of the minority spin channel at the Fe $3d_{z^2}$ orbital for the nanostructures shown in Figure 2.

| Structure | $n_{z^2}$ | MAE (meV/atom) |
|---|---|---|
| (a) | 0.029 | 0.89 |
| (b) | 0.030 | 0.76 |
| (c) | 0.223 | -2.01 |
| (d) | 0.196 | -2.00 |
| (e) | 0.028 | 1.07 |
| (f) | 0.031 | 0.70 |

**Figures**

FIG 1. (Color online) MAE of monolayer Fe atom as a function of the applied energy shift of the 3$d_{z^2}$ orbital. The inset red curve shows the PDOS of minority spin channel at the 3$d_{z^2}$ orbital when $V_{z^2} = -0.3$ eV. The shaded plot is the respective PDOS with $V=0$ for comparison. Green vertical dash line in the inset shows the Fermi level.

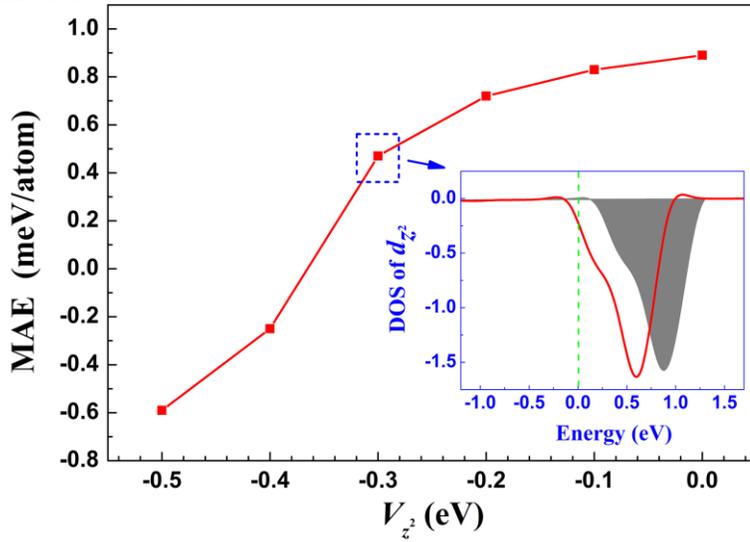

FIG 2. (Color online) Schematic of six Fe nanostructures. Here the nearest atom in-plane distance in (c) is $\sqrt{2}a$=4.06 Å, and other five structures have the nearest in-plane atom distance of $a$=2.87 Å ($a$ is the optimized lattice constant of Fe.).

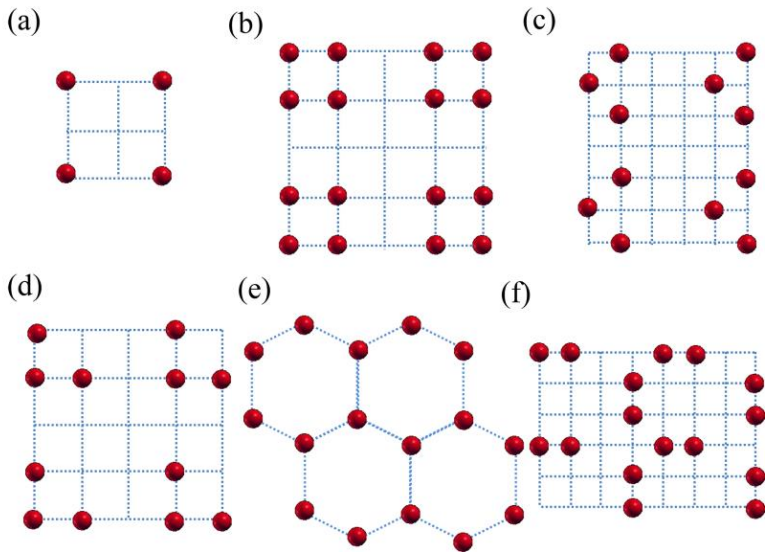

FIG 3. (Color online) (a) MAE of Fe/Cu film systems with different Fe nanostructures as a function of applied electric field. The black, red and blue scatters correspond to the Fe/Cu system with normal square, triangle and hexagon graphene-like Fe alignment respectively. The inset blue solid scatters show the dependence of MAE on the applied electric field in the normal square Fe alignment Fe/Cu system with an OSEP shift of -0.3 eV on the Fe 3 $d_{z^2}$ minority spin states. All the solid straight lines are fitted linearly to the calculated data. The MAE related magnetoelectric coefficient $β_s$ is calculated from the slope of every fitting line, in unit of $10^{-9}$ erg/V cm. (b) PDOS of $3d_{z^2}$ orbital of the Fe atom in different Fe/Cu nanostructures and the normal square Fe alignment Fe/Cu system with an OSEP shift of -0.3 eV on the Fe $3d_{z^2}$ orbital. Systems of the triangle and hexagon Fe alignment are shown as red dash line and blue straight line, respectively. The purple dash dot line and the shaded plots are the PDOS of the Fe atom in normal Fe alignment Fe/Cu system with and without an OSEP shift. The vertical dash line specifies the Fermi level.

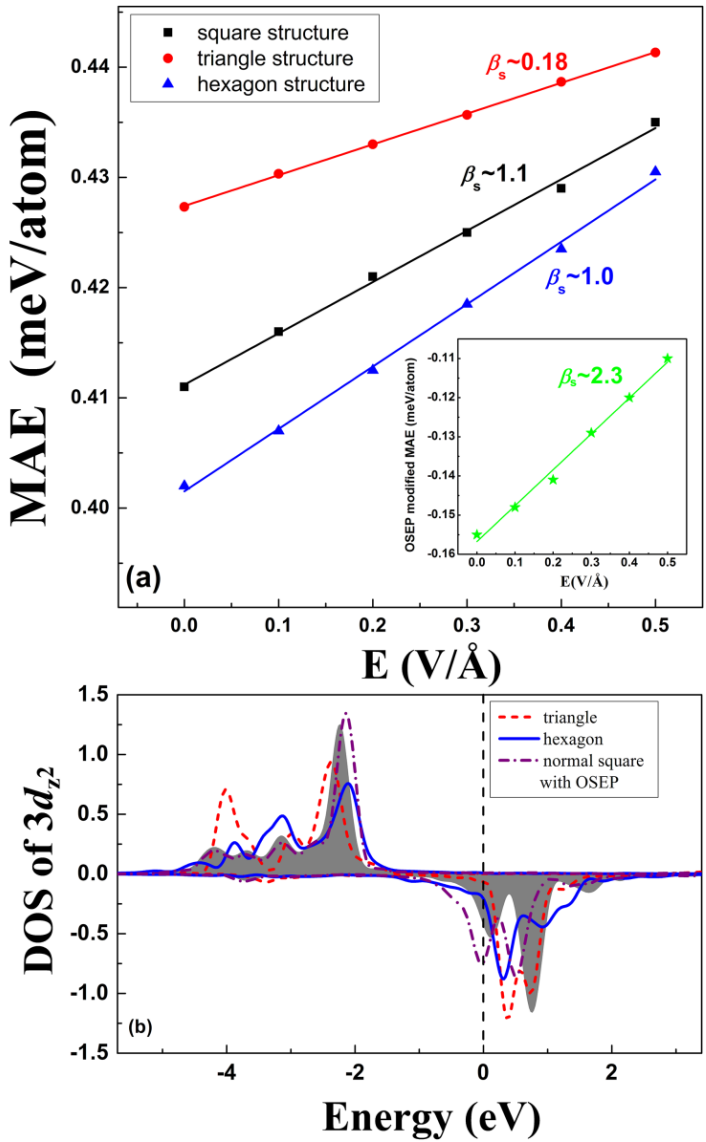